\documentclass[aps,floats,amssymb,twocolumn,superscriptaddress,showpacs]{revtex4}
\usepackage{calc}
\usepackage{psfrag}
\usepackage{graphicx}
\begin{document}
\def \beq{\begin{equation}}
\def \eeq{\end{equation}}

\begin{abstract}
We develop a non-perturbative approach to the quantum Hall bilayer
(QHB) at $\nu=1$ using trial wave functions.  We predict phases of
the QHB for arbitrary distance $d$ and, our approach, in a dual
picture, naturally introduces a new kind of quasiparticles - neutral
fermions. Neutral fermion is a composite of two merons of the same
vorticity and opposite charge.  For small $d$ (i.e. in the
superfluid phase), neutral fermions appear as dipoles. At larger $d$
dipoles dissociate into the phase of the two decoupled
Fermi-liquid-like states. This scenario is relevant for the
experimental situation where impurities  lock charged merons. In a
translation invariant (clean) system, continuous creation and
annihilation of meron-antimeron pairs  evolves the QHB toward a
paired phase.  The quantum fluctuations fix the form of the pairing
function to $g(z)=1/z^*$. A part of the description of the paired
phase is the 2D superconductor i.e. BF Chern-Simons theory. The
paired phase is not very distinct from the superfluid phase.
\end{abstract}


\title{Non-perturbative approach to the quantum Hall bilayer}

\author{M.V. Milovanovi\'{c}}
\affiliation{Institute of Physics, P.O.Box 68, 11080 Belgrade,
Serbia}

\author{Z. Papi\'{c}}
\affiliation{Institute of Physics, P.O.Box 68, 11080 Belgrade,
Serbia} \affiliation{Laboratoire de Physique des Solides,
Universit\'e Paris-Sud 11, 91405 Orsay, France}

\date{\today}
\maketitle \vskip2pc
\narrowtext

\section{Introduction}

The quantum Hall bilayer  at $\nu = 1$ consists of
two layers of two-dimensional electron gases  that are brought
close to one another in the quantum Hall regime of strong magnetic
fields. When the distance between the layers is much smaller than
the average distance between electrons inside each layer, inter and
intra Coulomb interactions are about the same. Then the expected
$\nu = 1$ state is the state of a single layer filled lowest Landau
level (LLL) generalized to two species. There is obvious degeneracy
in dividing electrons into two groups which leads to the phenomenon
of spontaneous symmetry breaking \cite{wz} and the existence of a
Goldstone mode \cite{sp1}. The expected superfluid behavior was
verified also by very large zero bias voltage peak in tunneling
conductance \cite{sp2}, but no clear evidence was found for finite
temperature Berezinskii-Kosterlitz-Thouless  (BKT) transition
\cite{moo} in transport experiments \cite{cfl}.

Therefore there is a need to systematically address the question of
superfluid disordering in the QHB. In particular there is a need to
understand the role of quantum disordering in this system that
becomes important as the distance between the layers is increased.
In most of the previous work the starting point for the
discussion of the physics of the bilayer was the ground state (GS)
for very small distance between the layers as a mean field
solution to which none or some corrections were developed
\cite{moo,lop}. We will take a non-perturbative approach inspired by
the Laughlin solution of the $\nu = 1/3$ problem in which we will
uniquely determine possible wave functions (WFs) for the GSs of the
bilayer at an arbitrary distance.

There are two basic paradigms of superfluid disordering that are known:
(1) BKT (2D XY model) for which the transition proceeds via
unbinding of dipoles of vortex-antivortex pairs, and (2) $\lambda$
transition type (3D XY model) for which the transition is
characterized by a condensation of vortex-antivortex loops
\cite{loopcon}.

On the other hand, in this paper, through an analysis of the allowed
possibilities for homogeneous WFs as the distance is varied, we will
identify two families of WFs and relate them to the two ways of disordering the QHB
superfluid mentioned previously. The families will be introduced in Section II.

One family, as it will turn out does not include elementary vortices
- merons of QHB, in its description of superfluid disordering.
Merons are part of the description of the QHB superfluid for small
distances as is well-known and well-established in Ref. \cite{moo}.
Therefore this family of (homogeneous) WFs can be relevant only for
dirty systems - systems with impurities, which can lock merons due
to merons being charged quasiparticles. Then the only vortices that
may participate in superfluid disordering and on which description
of this family of WFs is based, are neutral composites of two merons
of opposite charges - neutral vortices, and as we will find
fermionic quasiparticles that carry only layer degree of freedom. We
will show that the superfluid disordering of this family can be
understood through a Coulomb (fermionic) plasma picture of dipoles
of these neutral fermions. Therefore this family we can consider as
the one that exemplifies the BKT way of superfluid disordering, our
first paradigm. This whole picture will be corroborated by the fact
that the WFs of this family do not incorporate quantum fluctuations
(Section IV) and, therefore, do not incorporate quantum disordering
that is based on merons. The family from the viewpoint of a dual
description (i.e. in terms of quasiparticles - neutral fermions)
will be analyzed  in Section III.

The other family incorporates weak pairing among neutral fermions
and, as we will show, by assuming a special kind of pairing agrees
and correlates with the description of quantum fluctuations of the
usual superfluid disordering in a translatory invariant system that
one finds in other approaches (field-theoretical). It is expected
that this kind of disordering and pairing would lead to a charge
density wave (CDW) solution \cite{loy}. Still our general
considerations open possibilities for other kinds of weak pairing
that can be present in this quantum Hall system. The most likely
candidate is the one with pairing function $g(z) \sim
\frac{1}{z^{*}}$ that results in non-trivial corrections (from
quantum fluctuations and disordering) to the ground state wave
function as the distance is varied. In general we expect that a weak
pairing scenario will correspond to the superfluid disordering of
the usual superfluid in $2 + 1$ dimension and therefore to the class
of 3D XY, our second paradigm. The family with weak pairing ansatz
will be analyzed in Section IV.

With respect to the experiments, where impurities are necessarily
present and we can expect also inhomogeneous ground state solutions,
our homogeneous candidates of the first family (without  neutral
fermion weak pairing) are still possible solutions for which
transitions may proceed via dissociation of dipoles - pairs of
opposite vorticity neutral fermions. In this sense and as will be
more clear later, the quantum phase transitions with respect to
changing the distance in Refs. \cite{ch,kel,kar} correspond to this
dissociation. On the other hand, an analysis will show that in a
translatory invariant system meron excitations via their loop
condensation may produce an intercorrelated, paired liquid state
 for the neutral sector, if a transition to a CDW does not occur.

\section{Universality classes of ground states}
\subsection{Introduction}
A great deal is known
from the experimental and theoretical point of view of the QHB in
the two extremes when the distance between layers, $d$, is (1) much
smaller or (2) much larger than the magnetic length, $l_{B} =
(\hbar/e B)^{1/2}$, where $B$ is - the magnetic field, the characteristic
distance between the electrons inside any of the layers. When $d <<
l_{B}$, i.e. inter and intra Coulomb interactions are about the
same, the good starting point and description is so-called (111)
state \cite{halp},
\begin{equation}
\Psi_{111}(z_{\uparrow}, z_{\downarrow}) =
\prod_{i<j}(z_{i\uparrow}-z_{j\uparrow})
\prod_{k<l}(z_{k\downarrow}-z_{l\downarrow})\prod_{p,q}(z_{p\uparrow}-z_{q\downarrow})
\end{equation}
where $z_{i \uparrow}$ and $z_{i \downarrow}$ are two-dimensional
complex coordinates of electrons in upper and lower layer
respectively and we omitted the Gaussian factors. This is suggestive
of the exciton binding \cite{fermac}; any electron coordinate is
also zero of the WF for any other electron coordinate - the
correlation hole is just opposite to electron. This exciton
description can be a viewpoint of the phenomenon of superfluidity
found in these systems \cite{sp1,sp2} and is closely connected to
the concept of composite bosons (CBs) \cite{zh} that can be used as
natural quantum Hall quasiparticles in this system. When $d
>> l_{B}$ we have the case of the decoupled layers and the GS is a
product of single layer filling factor 1/2 WFs; each describes a
Fermi-liquid-like state \cite{rr},
\begin{equation}
\Psi_{1/2}(w) = {\cal P} \{ {\cal
 F}_{s}(w,\overline{w})\prod_{i<j}(w_{i\uparrow}-w_{j\uparrow})^2\}
\end{equation}
where ${\cal F}_{s}$ is the Slater determinant of free waves of
noninteracting particles in zero magnetic field and ${\cal P}$
represents projection to LLL. Underlying quasiparticles are
composite fermions (CFs), the usual quasiparticles of the single
layer quantum Hall physics.
\subsection{Two families - universality classes of wave functions}
To answer the question of intermediate distances we may try to,
classically speaking, divide electrons into two groups, one in which
electrons correlate as CBs and the other as CFs \cite{srm}. The
ratio between the numbers of CBs and CFs would be determined by the
distance between layers. The WF constructed in this way would need
an overall antisymmetrization in the end, but also intercorrelations
among the groups as each electron of the system sees the same number
of flux quanta through the system (equal to the number of
electrons). This requires that the highest power of any electron
coordinate is the same as the number of electrons in the
thermodynamic limit. If we denote by a line the Laughlin-Jastrow
factor $ \prod_{A,B} (z_{A} - z_{B})$ between two groups of
electrons, $A$ and $B$ ($A,B=CB,CF$), the possibilities for the QHB
GSWFs can be summarized as in Fig. 1.
\begin{figure}
\centering
\includegraphics[scale=0.35]{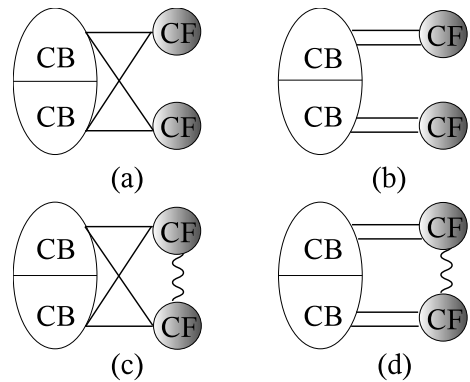}
\caption{Universality classes of wave functions}\label{unicl}
\end{figure}

If we ignore the possibility of pairing between CFs \cite{bon}
denoted by wriggly lines in Fig.1(c)and Fig.1(d) we have two basic families of
the GSWFs depicted in Fig.1(a) and Fig.1 (b). The requirement that each electron
sees the same number of flux quanta through the system equal to the
number of electrons (we are at $\nu = 1$) very much reduces the
number of possible states - wave functions in the mixed CB - CF
approach. We can consider, for example, the possibility $(a)$
depicted in Fig. 1 which stands for the following wave-function in
the LLL,
\begin{eqnarray}\label{wfa}
\nonumber
\Psi_1=\mathcal{P}\mathcal{A_{\uparrow}}\mathcal{A_{\downarrow}}\{\prod_{i<j}
(z_{i\uparrow}-z_{j\uparrow})
 \prod_{k<l} (z_{k\downarrow}-z_{l\downarrow}) \prod_{p,q}
 (z_{p\uparrow}-z_{q\downarrow})\\
\nonumber \times {\cal F}_s(w_\uparrow ,\overline{w}_\uparrow)
\prod_{i<j} (w_{i\uparrow}-w_{j\uparrow})^2 \\
\nonumber \times {\cal F}_s(w_\downarrow ,\overline{w}_\downarrow)
\prod_{k<l}
(w_{k\downarrow}-w_{l\downarrow})^2\\
\nonumber \times \prod_{i,j}(z_{i\uparrow}-w_{j\uparrow})
\prod_{k,l}(z_{k\uparrow}-w_{l\downarrow}) \\
\times \prod_{p,q}(z_{p,\downarrow}-w_{q,\uparrow})
\prod_{m,n}(z_{m\downarrow}-w_{n\downarrow})
 \}.
\end{eqnarray}
where ${\cal A}_{\uparrow}$ and ${\cal A}_{\downarrow}$ denote the
overall antisymmetrizations. In the thermodynamic limit, the
relation between the number of particles and flux quanta reads,
\begin{eqnarray}
N_{\phi}^{b} &=& N_{b \uparrow} + N_{b \downarrow} +  N_{f \uparrow}
+
N_{f \downarrow}, \nonumber \\
N_{\phi}^{f \uparrow} &=& 2 N_{f \uparrow}  +  N_{b \uparrow} +
N_{b \downarrow}, \nonumber \\
N_{\phi}^{f \downarrow} &=& 2 N_{f \downarrow}  +  N_{b \uparrow} +
N_{b \downarrow},
\end{eqnarray}
where we denoted by $N_{\phi}^{b}$  and $N_{\phi}^{f \sigma}$
separately the number of flux quanta that electrons that
correlate as CBs and CFs respectively see, $N_{b \sigma}$ and $N_{f
\sigma}$ are the number of CBs and CFs  inside the layer $\sigma$,
respectively  ($\sigma = \uparrow, \downarrow$ is the layer index).
The requirement constrains $N_{\phi} = N_{\phi}^{b} = N_{\phi}^{f
\sigma}$, where $N_{\phi}$ is the number of flux quanta through the
system. (This leads to the additional requirement $N_{f \uparrow} =
N_{f \downarrow}$ which leaves $N_{b \uparrow} - N_{b \downarrow}$
unconstrained, connected with the Bose condensation phenomenon that
the wave function should be part of (\cite{zpmvm,moo,fermac}).

 The only additional way to count the flux
quanta that electrons see, with the (symmetric  under $\uparrow
\leftrightarrow \downarrow$ reversal) application of the Jastrow -
Laughlin factors that we need to have, is:
\begin{eqnarray}
N_{\phi}^{b \uparrow} &=& N_{b \uparrow} + N_{b \downarrow} +  2
N_{f \uparrow},  \nonumber \\
N_{\phi}^{b \downarrow} &=& N_{b \uparrow} + N_{b \downarrow} +
2 N_{f \downarrow}, \nonumber \\
N_{\phi}^{f \uparrow} &=& 2 N_{f \uparrow}  +  2 N_{b \uparrow},  \nonumber \\
N_{\phi}^{f \downarrow} &=& 2 N_{f \downarrow}  +  2 N_{b
\downarrow},
\end{eqnarray}
which leads to the possibility $(b)$ (with constraints $N_{b
\uparrow} = N_{b \downarrow} $ and  $N_{f \uparrow} = N_{f
\downarrow}) $
 The intercorrelations in the first
family in Fig.1(a) are in the spirit of $\Psi_{111}$ correlations,
and those in the second family in Fig.1(b) are in the spirit of the
decoupled state, $\Psi_{1/2} \times \Psi_{1/2}$, where we correlate
exclusively inside each layer.
\subsection{Discussion}
We can imagine a mixture of both intercorrelations (of Fig. 1(a) and
Fig. 1(b)) in a single wave function  but these mixed states, by
their basic response \cite{zpmvm}, fall into one of the universality
classes depicted in Fig. 1. In Ref. \onlinecite{zpmvm} explicitly
such a mixture and possibility under name ``generalized vortex
metal" was considered, in the scope of a Chern - Simons (CS) theory, and
it was proved that it does not support a Goldstone (gapless) mode
which was found to exist for the state depicted in Fig. 1(a). These
generalized states belong to the universality class of the state
depicted in Fig. 1(b) for which in the scope of the same theory we
find in the low-energy spectrum only a gapped collective mode \cite{zpmvm}.

The Chern-Simons theory we mentioned neglects the overall
antisymmetrization built in the classes of Fig. 1. We can justify
this neglect (1) by taking a point of view that stems from similar
situations with quantum Hall states like hierarchy and Jain's
constructions that in the low-energy sector can be considered as
multicomponent systems \cite{nr} (we will argue later that the state
of Eq.(\ref{wfa}) can be mapped to a hierarchy construction), or (2)
a posteriori because the results of the effective description of the
classes in Fig. 1 are quite sensible and are expected for the states
 we are familiar with from numerics (the state in our Fig. 1(a) as analyzed in \cite{srm}). (We do
not ask this type of theory for detailed answers anyway.) In this
way it was found by us (Refs.\onlinecite{zpmvm} and
\onlinecite{mvm}) examining the basic response in the pseudospin
channel in the random phase approximation (RPA) of these Chern -
Simons theories that the states in Fig.1(a) and Fig.1(c) represent
superfluids, and the states in Fig.1(b) and Fig.1(d) represent disordered
superfluids, compressible and incompressible, respectively. (Later,
in a more complete study, we will find that the states of Fig. 1(d)
are also compressible in the neutral channel.)

The two basic possibilities of connecting two extremes as depicted
in Fig. 1, i.e. without and with pairing of CFs, must correspond to
the two possible ways or paradigms that we know of disordering a
superfluid. We will substantiate this claim further by examining the
two superfluid constructions (Fig. 1(a) and Fig. 1(c)) in more detail.

\section{Neutral fermions and BKT disordering}
\subsection{Dual picture of the first family of wave functions with
neutral fermions}

 Let us write out the unprojected in the LLL version
of the construction in Fig. 1(a)(Eq.(\ref{wfa})) in the following
way:
\begin{eqnarray}
 \Psi_{1}=\cal{A}_{\uparrow} \cal{A}_{\downarrow}&\{& \Psi_{111}(z_{\uparrow}, z_{\downarrow})
 \Psi_{1/2}(w_{\uparrow}) \Psi_{1/2}(w_{\downarrow}) \nonumber \\
&& \times
\prod_{i,j}(z_{i\uparrow}-w_{j\uparrow})\prod_{k,l}(z_{k\uparrow}-w_{l\downarrow}) \nonumber \\
&& \times
\prod_{p,q}(z_{i\downarrow}-w_{q\uparrow})\prod_{m,n}(z_{m\downarrow}-w_{n\downarrow})
  \},
\label{fir}
\end{eqnarray}
where, as before, $z_{\sigma}$'s  and $w_{\sigma}$'s denote
coordinates of
electrons belonging to the layer with index $\sigma$ and
$\cal{A}_{\uparrow}$ and $\cal{A}_{\downarrow}$, as before, stand
for the antisymmetrizations. Using ${\cal S}_{\uparrow}$ and
${\cal S}_{\downarrow}$, symmetrizers inside each layer, the same
function, $\Psi_{1}$, can be written as:
\begin{eqnarray}
\Psi_{1} &=& {\cal S}_{\uparrow} {\cal S}_{\downarrow}
\Big\{\frac{\prod_{k<l}(w_{k\uparrow}-w_{l\uparrow})\prod_{p<q}(w_{p\downarrow}-w_{q\downarrow})}
{\prod_{i,j}(w_{i\uparrow}-w_{j\downarrow})} \times \nonumber
\\& &\;\;\;\;\;\;\;\;\;\;{\cal F}_{s}(w_{\uparrow}) \times {\cal
F}_{s}(w_{\downarrow})\Big\} \Psi_{111}
\end{eqnarray}
where $\Psi_{111}$ denotes the Vandermonde determinant (Slater
determinant in the LLL) of all coordinates in which all groups
equally participate.

By using the expressions for the densities of electrons in each
layer, $ \rho^{\sigma}(\eta) = \sum_{i} \delta^{2} (\eta -
z_{i}^{\sigma})$, {\em here now $z_{\sigma}$'s denote all electrons
of the layer $\sigma$}, we can rewrite the wave function in the
following way,
\begin{eqnarray}
\Psi_{1}=
 \int d^{2}\eta_{1 \uparrow} \cdots \int d^{2}\eta_{n
\downarrow}
&&\frac{\prod_{k<l}(\eta_{k\uparrow}-\eta_{l\uparrow})\prod_{p<q}(\eta_{p\downarrow}-\eta_{q\downarrow})}
{\prod_{i,j}(\eta_{i\uparrow}-\eta_{j\downarrow})}\nonumber \\
&& {\cal F}_{s}(\eta_{\uparrow}) \times {\cal
F}_{s}(\eta_{\downarrow}) \times \nonumber \\
 &&\rho^{\uparrow}(\eta_{1 \uparrow}) \cdots
\rho^{\downarrow}(\eta_{n \downarrow}) \Psi_{111}(z_{\uparrow},
z_{\downarrow}), \label{sec} \nonumber \\
\end{eqnarray}
where $n$ is the total number of electrons that correlate as CFs.
The equality is exact; any time we have in the product of $\rho$'s the
same layer electron coordinate more than once, the Laughlin - Jastrow
factors of $\eta$'s in the same layer force the wave function to
become zero.
The expression in Eq.(\ref{sec}) reminds us of a dual picture in
terms of some quasiparticles with $\eta$ coordinates as in
\cite{lau}. Certainly we are not describing the incompressible
physics of a Laughlin state where quasihole operators of coherent
states span the basis of low-energy physics and allow the
description in terms of wave functions of quasiholes (dual picture)
\cite{lau}. Nevertheless we will argue that we can delineate a
sector (find a subspace) to which constructions (Eq.(\ref{sec})
where $n$ is arbitrary) belong, which is spanned by a quasiparticle
basis of some neutral fermionic quasiparticles.

To find those quasiparticles we will rewrite Eq.(\ref{sec}) as
\begin{eqnarray}
\Psi_{1}=
 \int d^{2}\eta_{1 \uparrow} \cdots \int d^{2}\eta_{n
\downarrow}
&&\frac{\prod_{k<l}|\eta_{k\uparrow}-\eta_{l\uparrow}|\prod_{p<q}|\eta_{p\downarrow}-\eta_{q\downarrow}|}
{\prod_{i,j}|\eta_{i\uparrow}-\eta_{j\downarrow}|}\nonumber \\
&& {\cal F}_{s}(\eta_{\uparrow}) \times {\cal
F}_{s}(\eta_{\downarrow}) \times \nonumber \\
 && \Big\{ \exp\{i \phi(\eta_{1 \uparrow} \cdots \eta_{n \downarrow})\} \times \nonumber \\
 && \rho^{\uparrow}(\eta_{1 \uparrow}) \cdots
\rho^{\downarrow}(\eta_{n \downarrow}) \Psi_{111}(z_{\uparrow},
z_{\downarrow}) \Big\} \label{thi} \nonumber \\
\end{eqnarray}
where  $\exp\{i \phi(\eta)\}$ factor denotes the phase part of the
Laughlin-Jastrow factors in front of the Fermi seas in
Eq.(\ref{sec}). With respect to Eq.(\ref{sec}) we are allowed to
take for definiteness  that the phase factor always vanishes when
any of two $\eta$'s (or more) from the same layer coincide.

Our first question may be why states as
\begin{equation}
\rho^{\uparrow}(\eta_{1 \uparrow}) \cdots \rho^{\downarrow}(\eta_{n
\downarrow}) \Psi_{111} \sim \Psi_{b}(\eta_{1 \uparrow},\ldots,
\eta_{n \downarrow})
\end{equation}
would not make a bosonic basis. We look for the following overlap:
\begin{equation}
\int dz_{1 \uparrow} \cdots \int dz_{N \downarrow} \Psi_{b}(\eta_{1
\uparrow}',\ldots, \eta_{n \downarrow}') \Psi_{b}(\eta_{1
\uparrow},\ldots, \eta_{n \downarrow}).
\end{equation}
In the expansion of the density sums we may get
\begin{equation}
\delta^{2}(\eta_{1}' -z_{1}^{\uparrow}) \delta^{2}(\eta_{2}'
-z_{1}^{\uparrow})\delta^{2}(\eta_{1} -z_{1}^{\uparrow})
\delta^{2}(\eta_{2} -z_{2}^{\uparrow}) \cdots
\end{equation}
which would lead to the following contribution after
$z$-integration:
\begin{eqnarray}
&&\delta^{2}(\eta_{1}' - \eta_{2}')\delta^{2}(\eta_{1} - \eta_{1}')
|\eta_{1} - \eta_{2}|^{2}
\exp\{-\frac{1}{2}(|\eta_{1}|^{2} + |\eta_{2}|^{2})\} \nonumber \\
&&\;\;\;\;\;\;\;\times \frac{1}{|\eta_{1} - \eta_{2}|^{2}}
\exp\{\frac{1}{2}(|\eta_{1}|^{2} + |\eta_{2}|^{2})\} \cdots
\label{bos}
\end{eqnarray}
The last term, before the dots, comes after the integration over
$z$'s that do not participate in the delta functions. As usual
\cite{lau} the term is the result of the screening of plasma which
we find in the plasma analogy of $\Psi_{111}$ state in its charge
channel. The term exactly cancels the preceding one (it is equal to
its inverse) and the same cancelation will happen for any pair of
$\eta$'s (in the place of $\cdots$) that in remaining
$z$-integration have the role of impurities (of charge one) in the
plasma of remaining $z$'s. This is very good because of our goal to
find basis states and leaves us to consider only delta functions in
the contribution. But we can see immediately in Eq.(\ref{bos}) that
$\delta(\eta_{1}' - \eta_{2}')$ spoils our goal that the states
mimic a Fock basis of bosonic particles. Therefore as candidates for
basis states we should consider fermionic states:
\begin{eqnarray}
|\eta_{1 \uparrow} \cdots \eta_{n \downarrow}> =
\frac{1}{\sqrt{n!\left(\begin{array}{c} N \\ n \end{array} \right)
}}\exp\{i \phi(\eta_{1 \uparrow} \cdots \eta_{n \downarrow})\} \times \nonumber \\
\rho^{\uparrow}(\eta_{1 \uparrow}) \cdots \rho^{\downarrow}(\eta_{n
\downarrow}) |\Psi_{111}> \label{sdef}
\end{eqnarray}
for which we can not get contributions of the type in Eq.(\ref{bos})
because the phase part does not allow two (or more) quasiparticles
to coincide. ( Eq.(\ref{sdef}) represents a fermionic state for
$\eta$ quasiparticles because of the phase part introduced in Eq.(9)
which is antisymmetric under the exchange of $\eta$'s.) Therefore we
should consider fermionic states in Eq.(\ref{sdef}) because of the
previously found non-desirable terms in the bosonic case (we are
looking for quasiparticles and their basis states that would have
features of the Fock space basis): the terms like the one with
$\delta(\eta^{'}_{1} - \eta^{'}_{2})$ in Eq.(\ref{bos}) lead to the
absence of orthogonality of these states which we would like to
represent coordinate basis states in the bosonic case and that can
be mended by taking fermions - then these terms are absent. By a
similar analysis which lead to Eq.(\ref{bos}), considering various
possibilities for delta function contributions of density operators
we can find that the leading - most singular and coherent behavior
of the states defined in Eq.(\ref{sdef}) is
\begin{eqnarray}
<\eta^{'}_{1 \uparrow}, \eta^{'}_{2 \uparrow} \cdots \eta^{'}_{n
\downarrow}|\eta_{1 \uparrow},\eta_{2 \uparrow} \cdots \eta_{n
\downarrow}>  \rightarrow \nonumber \\
\delta^{2}(\eta^{'}_{1 \uparrow} - \eta_{1 \uparrow})
\delta^{2}(\eta^{'}_{2 \uparrow} - \eta_{2 \uparrow}) \cdots
\delta^{2}(\eta^{'}_{n \downarrow} - \eta_{n \downarrow})- \nonumber \\
\delta^{2}(\eta^{'}_{1 \uparrow} - \eta_{2 \uparrow})
\delta^{2}(\eta^{'}_{2 \uparrow} - \eta_{1 \uparrow}) \cdots
\delta^{2}(\eta^{'}_{n \downarrow} - \eta_{n \downarrow})+ \cdots
\label{delnf}
\end{eqnarray}
The rest of contribution constitute incoherent phase factors with
fewer number ($ < n$) of delta functions but of the same kind as in
the leading behavior. We can not prove that the states make exactly
a Fock space of neutral fermionic quasiparticles, i.e. we do not
have an exact equality in Eq.(\ref{delnf}), but they stand fairly
close to that status. In other words we do not have the exact
equality in Eq.(15) i.e. equality to delta functions only
(appropriately antisymmetrized), but we have in addition some finite
contributions which can not change the fact that the overlap is
singular - at its maximum when $\eta^{'}$'s and $\eta$'s coincide.
Therefore quasiparticles are not point-like fermionic quasiparticles
(one certainly can not expect that from quasiparticles in a strongly
correlated system), they are extended, but clearly the overlap has
the singular contribution of antisymmetrized delta functions which
points out that we are fairly (to a good extent) close to the
fermionic Fock basis description. Even in the Laughlin case we can
not prove the exact LLL delta function overlaps of coherent states
of quasiholes. The quasiparticles are neutral because in the
construction of the states there is no net magnetic flux through the
system. See Eq.(\ref{sdef}) and the definition of the phase factor
in Eq.(\ref{thi}) with  Eq.(\ref{sec}).

Now that we know basis states, just by looking at Eq.(\ref{thi}) we
can read out the GSWF in the dual picture in terms of neutral
fermions,
\begin{equation}
\Psi_{dual}(\eta) =
\frac{\prod_{k<l}|\eta_{k\uparrow}-\eta_{l\uparrow}|\prod_{p<q}|\eta_{p\downarrow}-\eta_{q\downarrow}|}
{\prod_{i,j}|\eta_{i\uparrow}-\eta_{j\downarrow}|} {\cal
F}_{s}(\eta_{\uparrow}) {\cal F}_{s}(\eta_{\downarrow}) \label{fou}
\end{equation}
This is a wave function of a 2D Coulomb fermionic plasma. In the
literature 2D Coulomb fermionic plasma with same charge particles is
fairly known and explored \cite{barwen,assa}. It is a dynamical
system of fermionic particles in 2D that interact with the
long-range ($\sim - \ln\{r\}$) interaction. As shown in Ref.
\onlinecite{barwen} the Jastrow factor of the type $\prod_{i <j}
|z_{i} - z_{j}|^{\gamma}$ ($\gamma$ proportional to the interaction
coupling constant), together with multiplying Slater determinant of
free waves, describes the ground state function in the long-distance
limit. In our case we have a generalization of such a system to the
one with opposite charges. Assuming that the concentration of
particles is not large, which is the case of interest to us, we then
expect the dipole configurations of particles that the wave function
in Eq.(\ref{fou}) describes.

\subsection{Discussion}

Merons are true elementary vorticity quasiparticles of the
translatory invariant QHB system at least for small distances
between layers as shown in Ref. \cite{moo} and carry both charge and
vorticity. Therefore the neutral fermion basis that we described can
be a complete basis for the ground state evolution of the QHB in the
non-translatory invariant case in which merons by their charges are
bound to impurities.

 The wave function in Eq.(\ref{fou}) describes the superfluid
state in Fig 1(a). It encodes dipole positioning of opposite
vorticity (layer index) neutral fermions. With increasing distance
there are more dipoles of neutral fermions and they are expected to
be less tightly bound as in the description of a BKT disordering of
a 2D system with increasing temperature. Therefore we do not find
quantum fluctuations in this case. This will be explicitly shown by
calculations in the following section (see also Appendix A).

 In the superfluid phase,
with respect to merons, a neutral fermion dipole should be in
essence a superposition of quadrupolar combinations of merons - two
dipoles which come in pairs but at arbitrary distance as illustrated
in Fig. 2. In this way, as special configurations of dipoles,
neutral fermions, we expect, constitute the lowest lying states of
the QHB - (pseudo)spin or phonon waves \cite{sp1,zh}.
\begin{figure}
\centering
\includegraphics[width=0.8\linewidth]{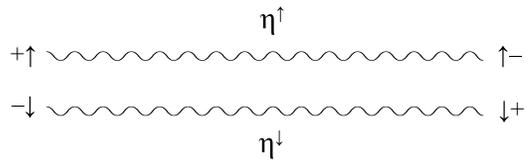}
\caption{The quadrupolar configurations of merons that make neutral
fermion pair. Compare the same configuration of Laughlin
quasiparticles as a description of ``magnetophonon" branch in the
Laughlin case in Ref. \onlinecite{zh}.}\label{quflu}
\end{figure}
If neutral fermions may be considered as eigenstates they must lie
very high in spectrum; like electrons in fractional quantum Hall
states they constitute the physics of $\Psi_{1}$ but their wave
function Eq.(\ref{fou}) describes a highly correlated state.

The dual expression of Eq.(\ref{fou}) was derived under assumption
of the screening properties in the charge channel of the particles
participating in the plasma analogy based on $\Psi_{111}$ state. As
the distance is increased there are less of them and the breakdown
of the description in terms of dipoles of neutral fermions at
smaller distances becomes a possibility. We expect that due to
impurities there will be patches (islands) of dissociated neutral
fermions \cite{sh}.

\section{Quantum fluctuations and quantum disordering}

\subsection{Introduction}
The two paradigms
- models of superfluid disordering as applied to our 2+1 dimensional
system mean that the time evolution is such that (1) meron
-antimeron pairs are locked on impurities  or (2) created and
annihilated at some later time and therefore making a loop in time.
 The loops in time signify the presence of quantum disordering \cite{loopcon}.
We will discuss and detect the presence of quantum disordering in
the WFs of class $(c)$ in Fig. 1 by examining how they relate to and
incorporate ordinary (not quantum disordering that involves merons -
vortices) quantum fluctuation phonon contribution in this case
\cite{lop,loy}. We will find that the WFs of class $(a)$ in Fig. 1
do not have this contribution.

\subsection{Quantum fluctuations due to phonons and quantum
disordering}

The usual \cite{zh} Chern-Simons (CS) field theory approach
\cite{loy} in the random phase approximation (RPA) to the bilayer
problem at $\nu = 1$ (which in the neutral channel reduces just to
the problem of ordinary superfluid with only phonon description and
contribution) finds the following correction to the $\Psi_{111}$
state:
\begin{equation}
\Psi_{PH} = \exp\Big\{ - \frac{1}{2} \sum_{k}
\frac{\sqrt{\frac{V_{-}(k)}{\rho_{E}}}}{k} \rho_{k}^{-}
\rho_{-k}^{-}\Big\} \Psi_{111}
\end{equation}
where $\rho_{k}^{-} = \rho_{k}^{\uparrow} - \rho_{k}^{\downarrow}$,
$V_{-}(k) = \frac{V_{\uparrow \uparrow}(k) - V_{\uparrow
\downarrow} (k)}{2}$, $V_{\uparrow \uparrow} = \frac{2 \pi}{k}$,
$V_{\uparrow \downarrow} = \frac{2 \pi}{k} \exp\{- k d\}$ i.e.
$V_{-}(k)$ is the interaction in the neutral channel, $\rho_{E} =
\frac{\bar{\rho}}{m}$ where $m$ is the electron mass and
$\bar{\rho}$ is the uniform total density. In the small $d$ limit
$V_{-}(k) = \pi d$ and we can expand the expression $\Psi_{PH}$ as
\begin{equation}
\Psi_{PH} = \Psi_{111} - (\sum_{k} \frac{c \sqrt{d}}{k} \rho_{-
k}^{-} \rho_{k}^{-}) \Psi_{111} + \cdots \label{dots}
\end{equation}
where $c$ is a positive constant. The terms after the first one
represent corrections, in the order of importance, to the
$\Psi_{111}$ ansatz as $d$ increases.

On the other hand the WFs of Fig. 1(c) are more general as they
suggest the form of the correction terms of wider class than the one
used in the expansion (Eq.(\ref{dots})) with only exception that the
class demands equal number of $\rho_{k}^{\uparrow}$'s and
$\rho_{k}^{\downarrow}$'s because in writing down the classes of
Fig. 1 we explicitly distinguished $\uparrow$'s from $\downarrow$'s
and fixed the number of $\uparrow$'s and $\downarrow$'s.

We can start comparing and relating the first phonon correction i.e.
\begin{equation}
\sim \sum_{k} \frac{1}{k} \rho_{- k}^{\uparrow}
\rho_{k}^{\downarrow}
\end{equation}
to a wave function of two neutral fermions ($\uparrow$ from
$\downarrow$) i.e. density operators as in Eq.(\ref{sec}) but with a
pairing between them as in class Fig. 1(c).

Without pairing we would have
\begin{equation}
\int d^{2}\eta_{1 \uparrow} \int d^{2}\eta_{2 \downarrow}
\frac{1}{(\eta_{1 \uparrow} - \eta_{2 \downarrow})}
\rho^{\uparrow}(\eta_{1 \uparrow}) \rho^{\downarrow}(\eta_{2
\downarrow}) \label{1+1}
\end{equation}
which is identical to zero (no correction) as can be found out in
Appendix A. This is an important result and shows that there are no
quantum fluctuations in the first family of WFs discussed in Section
III. Besides this analytical proof, our statement is further
corroborated by the fact that the computer-generated two neutral
fermion state also does not exist - see Ref. \cite{srm}.

 Therefore we continue by considering
\begin{equation}
\int d^{2}\eta_{1 \uparrow} \int d^{2}\eta_{2 \downarrow}
\frac{1}{|\eta_{1 \uparrow} - \eta_{2 \downarrow}|^{\alpha}}
\rho^{\uparrow}(\eta_{1 \uparrow}) \rho^{\downarrow}(\eta_{2
\downarrow})
\end{equation}
where $\alpha = 1$ if we take $g(z) = \sqrt{\frac{z}{z^{*}}}$ for
the pairing function or $ \alpha = 2$ if $g(z) = \frac{1}{z^{*}}$
For $\alpha = 1$ the expression in Eq.(\ref{1+1}) reduces to the
form of the first phonon contribution in the long-distance limit
with  the $\frac{1}{k}$ singularity (see Appendix A) and for $\alpha
= 2$ this singularity softens to $\sim - \ln\{k l_{B}\}$ where
$l_{B}$ is the magnetic length (see Appendix A). We will consider
only these most weakly pairing cases; the case $g(z) = \frac{1}{z}$
does not produce correction as can be seen in Appendix A.

Next we consider more than two density operator constructions i.e.
more than two neutral fermions constructions as in Eq.(\ref{sec})
but instead of the two decoupled Fermi seas we have a pairing
between neutral fermions:
\begin{eqnarray}
\Psi_{2}^{n} =  \int d^{2}\eta_{1 \uparrow} \cdots \int d^{2}\eta_{n
\downarrow}
&&\frac{\prod_{k<l}(\eta_{k\uparrow}-\eta_{l\uparrow})\prod_{p<q}(\eta_{p\downarrow}-\eta_{q\downarrow})}
{\prod_{i,j}(\eta_{i\uparrow}-\eta_{j\downarrow})}\nonumber \\
&& Det\{g(\eta_{\uparrow} - \eta_{\downarrow})\} \times \nonumber \\
 &&\rho^{\uparrow}(\eta_{1 \uparrow}) \cdots
\rho^{\downarrow}(\eta_{n \downarrow}) \Psi_{111}(z_{\uparrow},
z_{\downarrow})
\nonumber \\
= \int d^{2}\eta_{1 \uparrow} \cdots \int d^{2}\eta_{n \downarrow}&
& Det\{\frac{1}{\eta_{\uparrow} - \eta_{\downarrow}}\} \times \nonumber \\
&& Det\{\sqrt{\frac{\eta_{\uparrow} - \eta_{\downarrow}}{\eta_{\uparrow}^{*} - \eta_{\downarrow}^{*}}}\} \times \nonumber \\
 &&\rho^{\uparrow}(\eta_{1 \uparrow}) \cdots
\rho^{\downarrow}(\eta_{n \downarrow}) \Psi_{111}(z_{\uparrow},
z_{\downarrow}) \label{long}
\end{eqnarray}
where in the second expression we used the Cauchy determinant
identity i.e.
\begin{equation}
\frac{\prod_{k<l}(\eta_{k\uparrow}-\eta_{l\uparrow})\prod_{p<q}(\eta_{p\downarrow}-\eta_{q\downarrow})}
{\prod_{i,j}(\eta_{i\uparrow}-\eta_{j\downarrow})} =
Det\{\frac{1}{\eta_{\uparrow} - \eta_{\downarrow}}\}
\end{equation}
and substituted the pairing function that has lead us to the first
phonon correction for two paired neutral fermions. Immediately we
can see that the diagonal terms in which pairs of the two
determinants are the same would make further phonon-like corrections
i.e. their superposition with appropriate coefficients would lead to
\begin{equation}
\exp\{ - \sum_{k} \frac{c_{d}}{k} \rho_{ - k}^{\uparrow}
\rho_{k}^{\downarrow}\} \; \Psi_{111}. \label{phon}
\end{equation}
The other non-diagonal terms would lead to more complicated
constructions of four and more neutral fermions that should
participate in the description of quantum disordering i.e. describe
the physics beyond phonon contribution (\ref{phon}). Although in
some sense we are talking just about a class (a pool) of wave
functions that should describe quantum disordering we can fix
general form, at least for small $d$, of the superposition that
should completely model the ground state at fixed $d$
\begin{equation}
\Psi_{0} = \sum_{n = 0,2,\ldots} \Psi_{2}^{n}\; c_{n}. \label{nfcon}
\end{equation}
In the long distance limit (\ref{nfcon}) should tend to
(\ref{phon}). In other words non-diagonal terms in Eq.(\ref{long})
should be subleading to the leading behavior in (\ref{phon}). That
this is true from the physical point of view we expect that it is
enough to prove the subleading behavior in the case of four neutral
fermions $(\Psi_{2}^{4})$ and that can be found in Appendix B. The
proof is based on the smallness of higher order terms that may
appear inside the brackets in Eq.(\ref{phon}). This is assumed in
the  RPA approach and  expected in the  small $d$ limit.

Therefore the quantum Hall physics besides $g(z) =
\sqrt{\frac{z}{z^{*}}}$ pairing possibility brings or allows the
possibility of $g(z) = \frac{1}{z^{*}}$ pairing that introduces
non-trivial quantum corrections i.e. brings another kind of quantum
disordering. The $g(z) = \sqrt{\frac{z}{z^{*}}}$ accommodates the
usual (on the level of RPA) superfluid description in which we may
expect that the disordered phase will break translation symmetry.
Indeed, the bosonic CS field theories that are not based on quantum
Hall WFs give this scenario of the disordered phase as a charge
density wave \cite{loy}. It seems, therefore, there are two possible
scenarios for superfluid disordering not in the BKT class for the
bilayer in the translation symmetry invariant case (without
impurities). In the following we will discuss the second possibility
with $g(z) = \frac{1}{z^{*}}$ kind of pairing.

\subsection{Weak pairing $g(z) \sim \frac{1}{z^{*}}$ case and
conformal field theory considerations}

We expect, if the translational symmetry of the ground state remains
unbroken, that also in the case of pairing $g(z) = \frac{1}{z^{*}}$
the translatory invariant system smoothly evolves with the increase
of $d$ into the class of wave functions in Fig.1 (d). We would like
to know more about this class - whether it represents a distinct
phase.
 If we take  the choice $g(z) = \frac{1}{z^{*}}$
and examine the final form of the state of Fig.1(d) when there are
no CBs, we are lead to its following forms,
\begin{eqnarray}
\Psi_{2}& = & {\cal
D}et\{\frac{1}{z^{*}_{i\uparrow}-z^{*}_{j\downarrow}}\}
\prod_{i<j}(z_{i\uparrow}-z_{j\uparrow})^{2} \prod_{k<l}(z_{k
\downarrow}-z_{l \downarrow})^{2} \nonumber \\
& = &{\cal D}et\{\frac{1}{z^{*}_{i\uparrow}-z^{*}_{j\downarrow}}\}
{\cal D}et\{\frac{1}{z_{k \uparrow}-z_{l \downarrow}}\} \Psi_{111},
\end{eqnarray}
where to get the last line we used the Cauchy determinant identity.
The neutral part of $\Psi_{2}$ (not carrying a net flux through the
system as $\Psi_{111}$ does) that consists of the two determinants
can be viewed as a correlator of vertex operators of a single {\em
nonchiral} bosonic field. According to \cite{mr} conformal field
theory (CFT) correlators not only describe quantum Hall system WFs
but also can be used to find out about excitation spectrum and
connect to its edge and bulk theories. In this way motivated neutral
excitations are vertex operators that correspond to single-valued WF
expressions
 that multiply $\Psi_{2}$:
\begin{equation}
\exp\{i \delta_{1}  \phi(w,w^{*})\} \rightarrow \frac{\prod_{i}|z_{i
\uparrow} - w|^{2 \delta_{1}}}{\prod_{i}|z_{i \downarrow} - w|^{2
\delta_{1} }} \label{firstqp}
\end{equation}
\begin{equation}
\exp\{i \delta_{2}  \theta(w,w^{*})\} \rightarrow
\frac{\prod_{i}(z_{i \uparrow} - w)^{ \delta_{2}}}{\prod_{i}(z_{i
\uparrow}^{*} - w^{*})^{ \delta_{2}}} \frac{\prod_{i}(z_{i
\downarrow}^{*} - w^{*})^{ \delta_{2}}}{\prod_{i}(z_{i \downarrow} -
w)^{ \delta_{2}}}
\end{equation}
where $\phi(w,w^{*}) = \phi(w) + \phi(w^{*})$, $\theta(w,w^{*}) =
\phi(w) - \phi(w^{*})$, and $\phi(w)$ and $\phi(w^{*})$ are
holomorphic and antiholomorfic parts of the bosonic field
respectively. $\delta_{2}$ must be $\frac{1}{2}$ because of the
requirement of single valuedness. For detailed explanations of the
bosonic CFT analogies see Appendix C.

If the low-lying spectrum were consisting only of $\delta_{1
}=\frac{1}{2}$ and $\delta_{2}=\frac{1}{2}$ quasiparticle
excitations our system would be described by so-called BF
Chern-Simons theory or the theory of 2D superconductor \cite{han}.
The mutual statistics of quasiparticles - quasiparticles and
vortices, in this theory is semionic (due to the fact that vortices
carry half-flux $(\frac{h}{(2e)c})$ quantum) and that this is also
the case with our excitations can be easily checked via CFT
correlators - see Appendix C. Combining the analysis with the charge
part ($\Psi_{111}$) in which only charge 1 excitations are allowed
(half-flux quantum excitations are strongly confined \cite{gy}) we
may come to the conclusion that the degeneracy of the system GSs on
the torus must be 4 \cite{han,dem} But the expression for the first
kind of excitations (Eq.(\ref{firstqp})) allows a real continuum for
the value of $\delta_{1}$ exponent including $\delta_{1} = 0$ and
therefore we expect a compressible (gapless) behavior of the system
despite the incompressibility of the charge channel and seemingly
topological phase behavior in the neutral sector. Nevertheless we
expect that in our case BF CS theory is a part of the description of
the pairing phase in a Lagrangian in which there is a quadratic
non-derivative term in one of the two gauge fields; this allows a
branch of gapless excitations - see Appendix C for details.

The question may come why we did not do an analysis with the
projection to the LLL. Certainly the analysis is more involved where
``reversed flux part" i.e. complex conjugated determinant becomes an
operator that acts on the rest of wave function. Nevertheless, an
analysis of the edge excitation spectrum \cite{mj} suggests that it
can not conform to any description of simple, free CFT theories i.e.
can not belong to a totally incompressible class, and it is very
likely that the system is, as it follows from our unprojected
analysis, compressible in the neutral channel. Therefore it is very
hard to distinguish the physics of the states in Fig. 1(d) and Fig.
1(c) in the translatory invariant system that involve pairing of the
type $g(z) = \frac{1}{z^{*}}$.

 While we were finishing the writing a
numerical study (of homogenous WFs in the translatory invariant
case) appeared \cite{gsr} that agrees with and complements our
conclusions on the nature of pairing.

\section{Conclusions}
In conclusion, we presented two families of wave functions that
describe two possible ways of homogeneous disordering of the quantum
Hall superfluid with  their detailed description on the basis of the
dual (quasiparticle) picture of the quantum Hall effect. We also
presented detailed analysis of the disordering in the translation
invariant system on the basis of insights into the pairing function
of quasiparticles - neutral fermions. A class of candidate wave
functions was clearly connected with the formalism that we find in
other (Chern-Simons) theories, and the pairing function $ g(z) \sim
\frac{1}{z^{*}}$ was extracted as a clear choice that incorporates
quantum disordering and that will describe the system if it does not
transform into a CDW (charge density wave) inhomogeneous solution.

\section{Acknowledgments}
We thank A. Auerbach, G. M\"{o}ller,  E.H. Rezayi, S.H. Simon, Z.
Te\v{s}anovi\'{c} and especially N. Read for discussions.
M.V.M. gratefully acknowledges the hospitality of the Aspen Center
for Physics. The work was supported by Grant No. 141035 of the
Serbian Ministry of Science.
\appendix
\section{}
We want to prove
\begin{equation}
\int d^{2}\eta_{1 \uparrow} \int d^{2}\eta_{2 \downarrow}
\frac{1}{(\eta_{1 \uparrow} - \eta_{2 \downarrow})}
\rho^{\uparrow}(\eta_{1 \uparrow}) \rho^{\downarrow}(\eta_{2
\downarrow}) = 0
\end{equation}
After switching to the Fourier space, $\rho^{\sigma}(\eta) =
\sum_{k} \rho_{k}^{\sigma} \exp\{i \vec{k} \vec{\eta}\}$, the l.h.s.
becomes
\begin{equation}
2 \pi \sum_{k} \int d^{2}\eta \frac{1}{\eta} \exp\{i \vec{k}
\vec{\eta}\} \rho_{k}^{\uparrow} \rho_{- k}^{\downarrow} \label{f}
\end{equation}
The angle part of the integration with the help of the table
integral \cite{gr}:
\begin{equation}
\int_{0}^{\pi} \exp\{i \beta \cos x\} \cos\{ n x\} = i^{n} \pi
J_{n}(\beta) \label{ti1}
\end{equation}
yields
\begin{eqnarray}
\int d^{2} \eta \frac{1}{\eta} \exp\{i \vec{k} \vec{\eta}\} & = &
\int_{0}^{\infty} dr [i \pi J_{1}(k r) - i \pi J_{1}(- k r)]
\nonumber \\
& = & (-) \frac{i \pi}{k} \int_{0}^{\infty} dr \big[ \frac{d J_{0}(k
r)}{d r} + \frac{d J_{0}(- k r)}{d r}\big]\nonumber \\
& = & i \frac{2 \pi}{k}
\end{eqnarray}
where we used notation $|\vec{k}| = k$  and in the last line the
identity for the Bessel functions: $J_{0}^{'} = - J_{1}$. On the
other hand
\begin{equation}
\int d^{2} \eta \frac{1}{\eta} \exp\{- i \vec{k} \vec{\eta}\} = - i
\frac{2 \pi}{k},
\end{equation}
and therefore Eq.(\ref{f}) can be written as
\begin{eqnarray}
\pi \sum_{k} [\int d^{2}\eta \frac{1}{\eta} \exp\{i \vec{k}
\vec{\eta}\} \rho_{k}^{\uparrow} \rho_{- k}^{\downarrow} + \int
d^{2}\eta \frac{1}{\eta} \exp\{- i \vec{k} \vec{\eta}\} \rho_{-
k}^{\uparrow} \rho_{k}^{\downarrow} \nonumber \\
= i 2 \pi^{2} \sum_{k} \frac{1}{|\vec{k}|} (\rho_{k}^{\uparrow}
\rho_{- k}^{\downarrow} - \rho_{- k}^{\uparrow}
\rho_{k}^{\downarrow}) = 0   \;\;\;\;\;QED. \nonumber \\
\end{eqnarray}
Next we want to evaluate
\begin{equation}
\int d^{2}\eta_{1 \uparrow} \int d^{2}\eta_{2 \downarrow}
\frac{1}{|\eta_{1 \uparrow} - \eta_{2 \downarrow}|^{\alpha}}
\rho^{\uparrow}(\eta_{1 \uparrow}) \rho^{\downarrow}(\eta_{2
\downarrow})
\end{equation}
Again this reduces in the Fourier space to
\begin{equation}
2 \pi \sum_{k} \int d^{2}\eta \frac{1}{|\eta|^{\alpha}} \exp\{i
\vec{k} \vec{\eta}\} \rho_{k}^{\uparrow} \rho_{- k}^{\downarrow}
\label{ff}
\end{equation}
In the case of $\alpha = 1$ as usual for the real Coulomb
interaction in 2D the integral is
\begin{equation}
\int d^{2}\eta \frac{1}{|\eta|} \exp\{i \vec{k} \vec{r}\} = 2 \pi
\int_{0}^{\infty} dr J_{0}(k r) = \frac{2 \pi}{k}
\end{equation}
In the case of $\alpha = 2$ we have
\begin{equation}
\int d^{2}\eta \frac{1}{|\eta|^{2}} \exp\{i \vec{k} \vec{r}\} = 2
\pi \int_{0}^{\infty} dr \frac{J_{0}(k r)}{r} \label{int}
\end{equation}
The integral needs a cut-off at small distances (otherwise diverges)
which should be included in our effective description and as usual
can be taken to be $l_{B}$ (magnetic length distance). Therefore,
instead of Eq.(\ref{int}) we have
\begin{equation}
2 \pi \int_{0}^{\infty} dr \frac{r J_{0}(k r)}{r^{2} + l_{B}^{2}} =
2 \pi K_{0}(l_{B} k)
\end{equation}
In the small momentum limit we can approximate:
\begin{equation}
K_{0}(z) \approx - \ln\{\frac{z}{2}\} + o(z)
\end{equation}
and therefore our first phonon-like correction in this case of
pairing is
\begin{equation}
\sum_{k} (-) \ln\{k l_{B}\} \rho_{k}^{\uparrow} \rho_{-
k}^{\downarrow}.
\end{equation}
For the case of pairing $g(z) = \frac{1}{z}$ we have
\begin{equation}
\int d^{2}\eta_{1 \uparrow} \int d^{2}\eta_{2 \downarrow}
\frac{1}{(\eta_{1 \uparrow} - \eta_{2 \downarrow})^{2}}
\rho^{\uparrow}(\eta_{1 \uparrow}) \rho^{\downarrow}(\eta_{2
\downarrow})
\end{equation}
which reduces to the solving of the following Fourier transform
\begin{equation}
\int d^{2}\eta \frac{1}{\eta^{2}} \exp\{i \vec{k} \vec{\eta}\}
\end{equation}
With the help of Eq.(\ref{ti1}) we have for the value of the
integral:
\begin{equation}
\int_{0}^{\infty} dr \frac{1}{r} \{ - \pi J_{2}(k r) - \pi J_{2}(- k
r)\}
\end{equation}
We may use the table integral \cite{gr}
\begin{equation}
\int_{0}^{\infty} \frac{J_{\nu}(a x)}{x^{\nu - q}} dx =
\frac{\Gamma(\frac{1}{2} q + \frac{1}{2})}{2^{\nu - q} a^{q - \nu +
1}\Gamma(\nu - \frac{1}{2} q + \frac{1}{2})} \label{ti2}
\end{equation}
for $ - 1 < Re q < Re \nu - \frac{1}{2}$ to find out  that the value
of the integral does not depend on $k$ i.e.
\begin{equation}
\int d^{2}\eta \frac{1}{\eta^{2}} \exp\{i \vec{k} \vec{r}\} = - \pi
\end{equation}
Therefore the phonon-like correction in this case is proportional to
\begin{equation}
(\sum_{k} \rho_{k}^{\uparrow} \rho_{k}^{\downarrow}) \Psi_{111}
\end{equation}
and in the real (coordinate) space this becomes:
\begin{eqnarray}
&&\int d^{2} \eta \rho^{\uparrow}(\eta) \rho^{\downarrow}(\eta)
\Psi_{111} = \nonumber \\&& \sum_{i,j} \int d^{2} \eta
\delta^{2}(\eta - z_{i \uparrow}) \delta^{2}(\eta - z_{j
\downarrow}) \Psi_{111} = \nonumber \\ &&\sum_{i,j} \delta^{2}(z_{i
\uparrow} - z_{j \downarrow}) \Psi_{111} = 0
\end{eqnarray}
i.e. no correction at all.
\section{}
We consider non-diagonal (non-phonon-like) corrections that come
from the description of quantum disordering by the class of WFs in
Fig. 1(c) when the pairing is fixed to be $g(z) =
\sqrt{\frac{z}{z^{*}}}$, i.e. non-diagonal terms of Eq.(\ref{long})
with $n = 4$. We want to prove the subleading behavior with respect
to the diagonal terms as the one with $ \eta_{1 \uparrow} \equiv
\eta_{1}$,$ \eta_{3 \uparrow} \equiv \eta_{3}$,$ \eta_{2 \downarrow}
\equiv \eta_{2}$, and $ \eta_{4 \uparrow} \equiv \eta_{4}$ in
\begin{eqnarray}
\int d^{2} \eta_{1} \int d^{2} \eta_{3} \int d^{2} \eta_{2} \int
d^{2} \eta_{4} && \frac{1}{|\eta_{1} - \eta_{2}|}\frac{1}{|\eta_{3}
- \eta_{4}|} \times \nonumber \\
&&\rho^{\uparrow}(\eta_{1}) \rho^{\uparrow}(\eta_{3})
\rho^{\downarrow}(\eta_{2}) \rho^{\downarrow}(\eta_{4})\nonumber \\
&& = \big[\sum_{k} \frac{(2 \pi)^{2}}{k} \rho_{k}^{\uparrow} \rho_{-
k}^{\downarrow}\big]^{2} \nonumber \\  \label{ph4}
\end{eqnarray}
of the following non-diagonal term:
\begin{eqnarray}
\int d^{2} \eta_{1} \int d^{2} \eta_{3} \int d^{2} \eta_{2} \int
d^{2} \eta_{4} && \frac{1}{(\eta_{1} - \eta_{2})}\frac{1}{(\eta_{3}
- \eta_{4})} \times \nonumber \\
&&\sqrt{ \frac{(\eta_{1} - \eta_{4})}{(\eta_{1}^{*} - \eta_{4}^{*})}
\frac{(\eta_{3} - \eta_{2})}{(\eta_{3}^{*} -
\eta_{2}^{*})}}\times \nonumber \\
&&\rho^{\uparrow}(\eta_{1}) \rho^{\uparrow}(\eta_{3})
\rho^{\downarrow}(\eta_{2}) \rho^{\downarrow}(\eta_{4}).\nonumber \\
\label{nd}
\end{eqnarray}
The non-diagonal terms by their forms should describe different
processes from the phonon contributions i.e. from those as
$(\rho_{k_{1}}^{\uparrow} \rho_{- k_{1}}^{\downarrow}) \cdots
(\rho_{k_{\frac{n}{2}}}^{\uparrow} \rho_{-
k_{\frac{n}{2}}}^{\downarrow})$ for arbitrary $k$'s. In the
long-distance approximation we will argue that the non-diagonal term
(Eq.(\ref{nd})) carry less importance that the phonon contribution
with the same number of density operators.

Introducing $\eta = \eta_{1} - \eta_{4}, \tilde{\eta} = \eta_{3} -
\eta_{2}, \eta_{-} = \eta_{1} - \eta_{3},$ and $\eta_{+} = \eta_{1}
+ \eta_{3}$ we can rewrite Eq.(\ref{nd}) as
\begin{eqnarray}
\sum_{k_{1},k_{3},\tilde{k},k}\int d^{2} \eta \int d^{2}
\tilde{\eta} &&\int d^{2} \eta_{-} \int d^{2} \eta_{+}
\frac{1}{(\eta_{-} + \tilde{\eta})}\frac{1}{(\eta
- \eta_{-})}\nonumber \\
&&\times\sqrt{\frac{\eta \;\tilde{\eta}}{\eta^{*} \tilde{\eta}^{*}}}
 \exp\{i \vec{\eta_{1}} \vec{k}_{1}\}
 \exp\{i \vec{\eta_{3}} \vec{k}_{3}\}\nonumber \\
&&\times\exp\{i (\frac{1}{2} \vec{\eta}_{+} - \frac{1}{2} \vec{\eta}_{-} - \vec{\tilde{\eta}}) \vec{\tilde{k}}\}\nonumber \\
 &&\times\exp\{i (\frac{1}{2} \vec{\eta}_{+} + \frac{1}{2} \vec{\eta}_{-} - \vec{\eta}) \vec{k}\} \nonumber \\
&&\times \rho^{\uparrow}_{k_{1}} \; \rho^{\uparrow}_{k_{3}} \;
\rho^{\downarrow}_{\tilde{k}} \; \rho^{\downarrow}_{k} \label{ndd}
\end{eqnarray}
The $\eta_{+}$ integration brings the constraint $ \vec{k} +
\vec{\tilde{k}} + \vec{k}_{1} + \vec{k}_{3} = 0$. Then the remaining
$\eta_{-}$ integration gives the following contribution,
\begin{eqnarray}
\int d^{2} \eta_{-} \frac{1}{(\eta_{-} +
\tilde{\eta})}\frac{1}{(\eta - \eta_{-})} \exp\{i
\frac{\vec{\eta}_{-}}{2} (\vec{k}_{1} - \vec{k}_{3} -
\vec{\tilde{k}} + \vec{k})\} = \nonumber \\
- i \frac{2 \pi}{|\vec{k} + \vec{k}_{3}|} \frac{1}{\eta +
\tilde{\eta}} [ \exp\{ i \vec{\tilde{\eta}} (\vec{k}_{3} +
\vec{\tilde{k}})\} - \exp\{ - i \vec{\eta} (\vec{k}_{3} +
\vec{\tilde{k}})\} ] \nonumber \\
\end{eqnarray}
where we used the constraint. Therefore the contribution is
proportional to
\begin{eqnarray}
&& \sum_{k_{3},\tilde{k},k}\frac{1}{|\vec{k} + \vec{k}_{3}|}\int
d^{2} \eta \int d^{2} \tilde{\eta} \frac{1}{(\eta +
\tilde{\eta})}\sqrt{\frac{\eta \;\tilde{\eta}}{\eta^{*} \tilde{\eta}^{*}}}\times\nonumber \\
&&\;\;\;\;\;[ \exp\{ i \vec{\tilde{\eta}} (\vec{k}_{3} +
\vec{\tilde{k}})\} - \exp\{- i \vec{\eta} (\vec{k}_{3} +
\vec{\tilde{k}})\} ]
 \nonumber \\
&&\;\;\;\;\;\;\;\times\exp\{- i  \tilde{\eta} \tilde{k}\}\exp\{- i  \eta k\}\nonumber \\
&&\;\;\;\;\;\;\;\times \rho^{\uparrow}_{- \tilde{k} - k - k_{3}} \;
\rho^{\uparrow}_{k_{3}} \; \rho^{\downarrow}_{\tilde{k}} \;
\rho^{\downarrow}_{k} \label{nndd}
\end{eqnarray}
In the long-distance limit $|\vec{k} + \vec{k}_{3}| \rightarrow 0$
but that does not cancel the part of the 2D volume in the
integration measure like in the phonon contribution (that would damp
the contribution) but is canceled by the difference of the
exponentials in the same limit in Eq.(\ref{nndd}). There is only one
more factor i.e. $\frac{1}{\eta + \tilde{\eta}}$ that can bring the
momentum inverse contribution but this only enforces $ k \approx
\tilde{k}$ i.e. $(\rho_{k}^{\uparrow} \rho_{- k}^{\downarrow})^{2}$
without a significant coefficient. This will only give the next
order contribution inside the brackets in Eq.(\ref{phon}) which for
small $d$, and as usual in the RPA  approach, we can neglect.
\section{}
We will give a more general view of the CFT analogies of so-called
\cite{fre} doubled CS theories to which BF CS theory belongs. In the
work of Freedman et al. (\cite{fre}) BF CS theory was classified as
the low-energy theory of the deconfined phase of $\mathbb{Z}_{2}$
gauge theory. There also $SU(2)_{1}\times\overline{SU(2)}_{1}$
doubled CS theory was considered. For the detailed description of
these theories the reader should consult Refs. \onlinecite{fre} and
\onlinecite{han}. Here we will, by writing down relevant CFT
correlators, demonstrate the analogies between non-chiral - complete
CFTs and these doubled CS theories.

First we will consider $SU(2)_{1}\times\overline{SU(2)}_{1}$ case.
The possible wave function with coordinates of two species $z_{1
\uparrow}, \ldots, z_{N \downarrow}$, for which there are equal
number of $\uparrow$'s and $\downarrow$'s: $N_{\uparrow} =
N_{\downarrow}$ and $N_{\uparrow} + N_{\downarrow} = N$, is
\begin{eqnarray}
 \Psi  & =&
\frac{\prod_{k<l}|z_{k\uparrow}-z_{l\uparrow}|
\prod_{p<q}|z_{p\downarrow}-z_{q\downarrow}|}
{\prod_{i,j}|z_{i\uparrow}-z_{j\downarrow}|}\nonumber \\
& = &\frac{\prod_{k<l}  \sqrt{z_{k\uparrow}-z_{l\uparrow}}
\prod_{p<q} \sqrt{z_{p\downarrow}-z_{q\downarrow}}}
{\prod_{i,j}\sqrt{z_{i\uparrow}-z_{j\downarrow}|}} \times \nonumber
\\
&& \frac{\prod_{k<l}\sqrt{z_{k\uparrow}^{*}-z_{l\uparrow}^{*}}
\prod_{p<q}\sqrt{z_{p\downarrow}^{*}-z_{q\downarrow}^{*}}}
{\prod_{i,j}\sqrt{z_{i\uparrow}^{*}-z_{j\downarrow}^{*}}}
\label{psi}
\end{eqnarray}
We use the following correlator of  vertex operators of a bosonic
field $\phi$:
\begin{equation}
<\exp\{i \beta  \phi(z_{1},z_{1}^{*})\} \exp\{- i \beta
\phi(z_{2},z_{2}^{*})\}> = \frac{1}{|z_{1} - z_{2}|^{2 \beta^{2}}}
\end{equation}
If $\alpha = \frac{1}{\sqrt{2}}$ we can rewrite our wave function as
\begin{eqnarray}
\Psi = <\exp\{i \alpha  \phi(z_{1},z_{1}^{*})\} \exp\{i \alpha
\phi(z_{2},z_{2}^{*})\} \dots \nonumber \\ \exp\{- i \alpha
\phi(z_{N},z_{N}^{*})\}>
\end{eqnarray}
and define:
\begin{equation}
\phi(z,z^{*}) = \phi(z) + \phi(z^{*})
\end{equation}
\begin{equation}
\theta(z,z^{*}) = \phi(z) - \phi(z^{*})
\end{equation}
Inserting a neutral pair $(w_{1}$ and $ w_{2})$ of $\exp\{i
\delta_{1} \phi(w, w^{*})\}$ vertex operators or $\exp\{i \delta_{2}
\phi(w, w^{*})\}$ vertex operators we can conclude that these
insertions correspond to multiplying the wave function $\Psi$
(Eq.(\ref{psi})) by:
\begin{equation}
\exp\{i \delta_{1}  \phi(w,w^{*})\} \rightarrow \frac{\prod_{i}|z_{i
\uparrow} - w|^{2 \delta_{1} \alpha}}{\prod_{i}|z_{i \downarrow} -
w|^{2 \delta_{1} \alpha}} \label{qp1}
\end{equation}
\begin{equation}
\exp\{i \delta_{2}  \theta(w,w^{*})\} \rightarrow
\frac{\prod_{i}(z_{i \uparrow} - w)^{ \delta_{2}
\alpha}}{\prod_{i}(z_{i \uparrow}^{*} - w^{*})^{ \delta_{2} \alpha}}
\frac{\prod_{i}(z_{i \downarrow}^{*} - w^{*})^{ \delta_{2}
\alpha}}{\prod_{i}(z_{i \downarrow} - w)^{ \delta_{2}
\alpha}}\label{qp2}
\end{equation}
(The general formula for the many vertex correlator can be found,
for example, in Ref. \onlinecite{ts}.) The single-valuedness of the
WFs demands $\delta_{2} = \frac{1}{\sqrt{2}}$. If we take also
$\delta_{1} = \frac{1}{\sqrt{2}} = \delta_{2} = \delta$ then
\begin{eqnarray}
<\exp\{i \delta  \phi(w_{1},w_{1}^{*})\} \exp\{- i \delta
\phi(w_{2},w_{2}^{*})\} \times \nonumber \\ \exp\{i \delta
\theta(w_{3},w_{3}^{*})\} \exp\{- i \delta
\theta(w_{4},w_{4}^{*})\}> = \nonumber \\
=\frac{1}{|w_{1} - w_{2}|^{2 \delta^{2}}} \frac{1}{|w_{3} -
w_{4}|^{2 \delta^{2}}} \times \nonumber \\\frac{(w_{1} - w_{3})^{
\delta^{2}}}{(w_{1}^{*} - w_{3}^{*})^{ \delta^{2}}} \frac{(w_{2} -
w_{4})^{ \delta^{2}}}{(w_{2}^{*} - w_{4}^{*})^{ \delta^{2}}}\times \nonumber \\
\frac{(w_{1}^{*} - w_{4}^{*})^{ \delta^{2}}}{(w_{1} - w_{4})^{
\delta^{2}}} \frac{(w_{2}^{*} - w_{3}^{*})^{ \delta^{2}}}{(w_{2} -
w_{3})^{ \delta^{2}}}\;\;\;
\end{eqnarray}
and the mutual statistics between any of two particles of different
kinds ((13),(14),(23), or (24)) is fermionic. To see that, for
example, for the (13) pair we send 2 towards 4 and switch $w_{1}$
and $w_{3}$ coordinates.

In our case of the quantum Hall bilayer,
\begin{eqnarray}
\Psi'& = & \frac{\prod_{k<l}|z_{k\uparrow}-z_{l\uparrow}|^{2}
\prod_{p<q}|z_{p\downarrow}-z_{q\downarrow}|^{2}}
{\prod_{i,j}|z_{i\uparrow}-z_{j\downarrow}|^{2}} \nonumber \\
& = & Det\{\frac{1}{z^{*}_{i\uparrow}-z^{*}_{j\downarrow}}\}
Det\{\frac{1}{z_{k \uparrow}-z_{l \downarrow}}\} ,
\end{eqnarray}
The same analysis as above will fix $\alpha = 1$ and $\delta =
\frac{1}{2}$ so that in this case the mutual statistics is semionic
just as it should be in the BF CS field theory.

The BF CS theory of a 2D superconductor is \cite{han}
\begin{equation}
\frac{1}{\pi} \epsilon^{\mu \nu \lambda} b_{\mu} \partial_{\nu}
a_{\lambda} - a_{\mu} j^{\mu} - b_{\mu} \tilde{j}^{\mu}
\end{equation}
where $a^{\mu}$ and $b^{\mu}, \mu = 0, 1, 2$ are gauge fields; the
first term is the CS term and $j^{\mu}$ and $\tilde{j}^{\mu}, \mu =
0, 1, 2$ represent quasiparticle and vortex density-currents. The
Lagrangian encodes in the $\frac{1}{\pi}$ coefficient mutual
semionic statistics between the two excitations in 2D superconductor
- any time quasiparticle encircles vortex it gets  the Bohm -
Aharonov phase $\pi$, because vortex corresponds to the half-flux
quantum excitation in the paired system. Higher order in derivatives
i.e. Maxwell terms $\sim (\partial \times a)^{2}$ and $\sim
(\partial \times b)^{2}$ are present in the description of the
ordinary ($s$ - wave) - gapped 2D superconductor and can describe
the plasmon modes that are gapped - see Ref. \onlinecite{han}. In
our case, because from CFT analogies (Eq.(\ref{qp1}) and
Eq.(\ref{qp2})) we find that $\delta_{1}$ can be continuous and
correspond to a branch of gapless excitations, we expect a quadratic
in one of the gauge fields, without derivatives, term to describe
such a behavior. For example, if we add a term quadratic in $b$
($b_{\mu} b^{\mu}$ with the $(\partial \times a)^{2}$ Maxwell term
present) our classical equations of motion will be : $ \Delta a = 0$
and $\partial \times b = 0$. They describe gapless behavior
(Goldstone mode) in one gauge field and associated quasiparticle
description, and incompressible behavior in the other.

(The $SU(2)_{1}\times\overline{SU(2)}_{1}$ theory can be described
by the following Lagrangian
\begin{equation}
\frac{1}{2 \pi} \epsilon^{\mu \nu \lambda} b_{\mu} \partial_{\nu}
a_{\lambda} - a_{\mu} j^{\mu} - b_{\mu} \tilde{j}^{\mu}
\end{equation}
and we see explicitly mutual fermionic statistics.)


\begin{references}
\bibitem{wz} X.-G. Wen and A. Zee, Phys. Rev. Lett. {\bf 69}, 1811 (1992).
\bibitem{sp1} I.B. Spielman,
J.P. Eisenstein, L.N. Pfeiffer, and K.W. West, Phys. Rev. Lett. {\bf
84}, 5808 (2000).
\bibitem{sp2} I.B. Spielman, J.P. Eisenstein, L.N. Pfeiffer, and K.W. West, Phys. Rev. Lett.
{\bf 87}, 036803 (2001).
\bibitem{moo} K. Moon et al., Phys. Rev. B {\bf 51}, 5138 (1995).
H. Mori, Kun Yang, S.M. Girvin, A.H. MacDonald, L. Zheng, D.
Yoshioka, and S.-C. Zhang, Phys. Rev. B {\bf 51}, 5138 (1995).
\bibitem{cfl} M. Kellogg, J.P. Eisenstein, L.N. Pfeiffer, and K.W. West, Phys. Rev. Lett.
{\bf 93}, 036801 (2004).
\bibitem{lop} A. Lopez and E. Fradkin, Phys. Rev. B {\bf 51}, 4347 (1995).
\bibitem{loopcon} L. Onsager, Nouvo Cim. {\bf 6}, 279
(1949); R.P. Feynman, in {\em Progress in Low Temperature Physics},
edited by C.J. Gorter, North-Holland, Amsterdam (1955); G. Williams,
Phys. Rev. Lett. {\bf 59}, 1926 (1987); B. Chattopadhyay and S.R.
Shenoy, Phys. Rev. Lett. {\bf 72}, 400 (1994); W. Janke and T.
Matsui, Phys. Rev. B {\bf 42}, 10 673 (1990); A.K. Nguyen and A.
Sudbo, Phys. Rev. B {\bf 57}, 3123 (1998).
\bibitem{loy} L. Jiang and J. Ye, Phys. Rev. B {\bf 74}, 245311
(2006); J. Ye and L. Jiang, Phys. Rev. Lett. {\bf 98}, 236802
(2007).
\bibitem{ch} A.R. Champagne, J.P. Eisenstein, L.N. Pfeiffer, and K.W. West, Phys. Rev. Lett.
{\bf 100}, 096801 (2008).
\bibitem{kel}M. Kellogg, J.P. Eisenstein, L.N. Pfeiffer, and K.W. West, Phys. Rev. Lett.
{\bf 90}, 246801 (2003).
\bibitem{kar} B. Karmakar, V. Pellegrini, A. Pinczuk, L.N. Pfeiffer,
and K.W. West, Phys. Rev. Lett. {\bf 102}, 036802 (2009).
\bibitem{halp} B.I.Halperin, Helv. Phys. Acta {\bf 56}, 75 (1983).
\bibitem{fermac} H. Fertig, Phys. Rev. B {\bf 40}, 1087 (1989);
 A.H. MacDonald, Physica B 298, 129 (2001).
\bibitem{zh}S.-C. Zhang, Int. J. of Mod. Phys. B {\bf 6}, 25
(1992).
\bibitem{rr} E.H. Rezayi and N. Read, Phys. Rev. Lett. {\bf 72}, 900
(1994).
\bibitem{srm} S.H. Simon, E.H. Rezayi, and M.V. Milovanovi\'{c}, Phys. Rev. Lett.
{\bf 91}, 046803 (2003).
\bibitem{bon}  N.E. Bonesteel, Phys. Rev. B {\bf 48}, 11484 (1993).
\bibitem{zpmvm} Z. Papi\'{c} and M.V. Milovanovi\'{c}, Phys. Rev. B {\bf 75}, 195304 (2007).
\bibitem{nr} N. Read, Phys. Rev. Lett. 65, 1502 (1990); N. Regnault,
M.O. Goerbig, and Th. Jolic\oe ur, Phys. Rev. Lett. {\bf 101},
066803 (2008); R. de Gail, N. Regnault, and M.O. Goerbig, Phys. Rev.
B {\bf 77}, 165310 (2008).
\bibitem{mvm} M.V. Milovanovi\'c, Phys. Rev. B {\bf 75}, 035314
(2007).
\bibitem{lau} R.B. Laughlin, in {\em The Quantum Hall Effect}, 2nd.
ed., edited by R.E. Prange and S.M. Girvin (Springer, New York,
1990).
\bibitem{barwen}P.-A. Bares and X.-G. Wen, Phys. Rev. B {\bf 48}, 8636 (1993).
\bibitem{assa} N. Lindner, A. Auerbach and D.Arovas,
cond-mat/0701571.
\bibitem{sh} A. Stern and B.I. Halperin, Phys. Rev. Lett.{\bf 88}, 106801 (2002).
\bibitem{mr} G. Moore and N. Read, Nucl. Phys. B {\bf 360}, 362
(1991); for the non-chiral arguments see M. Freedman, C. Nayak, K.
Shtengel, K. Walker, and Z. Wang, Ann. Phys. {\bf 310}, 428 (2004).
\bibitem{han} T.H. Hansson, V. Oganesyan, and S.L. Sondhi, Ann.
Phys. {\bf 313}, 497 (2004).
\bibitem{gy} G.S. Jeon and J. Ye, Phys. Rev. B {\bf 71}, 125314
(2005 ).
\bibitem{dem} E. Demler, C. Nayak, and S. Das Sarma, Phys. Rev. Lett. 86, 1853
(2001); Y.B. Kim et al., Phys. Rev. B {\bf 63}, 205315 (2001).
\bibitem{mj} M.V. Milovanovi\'c and Th. Jolic\oe ur,
arXiv:0812.3764.
\bibitem{gsr} G. M\"{o}ller, S.H. Simon, and E.H. Rezayi,
Phys. Rev. Lett. {\bf 101}, 176803 (2008); G. M\"{o}ller, S.H.
Simon, and E.H. Rezayi, Phys. Rev. B 79, 125106 (2009).
\bibitem{gr} I.S. Gradstein and I.M. Ryzhik, Tables, Fizmatgiz (in
Russian), Moscow (1962).
\bibitem{fre} M. Freedman, C. Nayak, K. Shtengel, K. Walker, and Z. Wang, Ann. Phys.
{\bf 310}, 428 (2004).
\bibitem{ts} C. Itzykson and J.-M. Drouffe, {\em Statistical Field
Theory}, Cambridge University Press, Cambridge (1991).
\end{references}
\end{document}